\documentclass[%
 reprint,
 amsmath,amssymb,
 aps,
]{revtex4-2}

\usepackage{graphicx}
\usepackage{dcolumn}
\usepackage{bm}
\usepackage{times}

\usepackage{siunitx}
\usepackage{hyperref}
\usepackage{comment}
\usepackage{amsmath,amssymb}
\usepackage[english]{babel}

\begin{document}

\preprint{APS/123-QED}

\title{Ultra-sensitive separation estimation of optical sources}

\author
{Clémentine Rouvière,$^{1\ast}$ David Barral,$^{1}$ Antonin Grateau,$^{1}$ Ilya Karuseichyk,$^{1}$ Giacomo Sorelli,$^{2}$ Mattia Walschaers,$^{1}$ Nicolas Treps$^{1}$\\
\textit{$^{1}$Laboratoire Kastler Brossel, Sorbonne Universit\'{e}, CNRS, ENS-Universit\'{e} PSL,  Coll\`{e}ge de France,}
\textit{4 place Jussieu, Paris, F-75252, France}\\
\textit{$^{2}$Fraunhofer IOSB, Ettlingen, Fraunhofer Institute of Optronics,}
\textit{System Technologies and Image Exploitation, Gutleuthausstr. 1, 76275 Ettlingen, Germany}\\
\normalsize{$^\ast$E-mail:  clementine.rouviere@lkb.upmc.fr.}
}

\date{\today}

\begin{abstract}
Historically, the resolution of optical imaging systems was dictated by diffraction, and the Rayleigh criterion was long considered an unsurpassable limit. 
In superresolution microscopy, this limit is overcome by manipulating the emission properties of the object. However, in passive imaging, when sources are uncontrolled, reaching sub-Rayleigh resolution remains a challenge.  
Here, we implement a quantum-metrolgy-inspired approach for estimating the separation between two incoherent sources, achieving a sensitivity five orders of magnitude beyond the Rayleigh limit. 
Using a spatial mode demultiplexer, we examine scenes with bright and faint sources, through intensity measurements in the Hermite-Gauss basis. 
Analysing sensitivity and accuracy over an extensive range of separations, we demonstrate the remarkable effectiveness of demultiplexing for sub-Rayleigh separation estimation. These results effectively render the Rayleigh limit obsolete for passive imaging.
\end{abstract}

\maketitle

\section*{Introduction}
\begin{figure}
    \centering
    \includegraphics[width=0.45\textwidth]{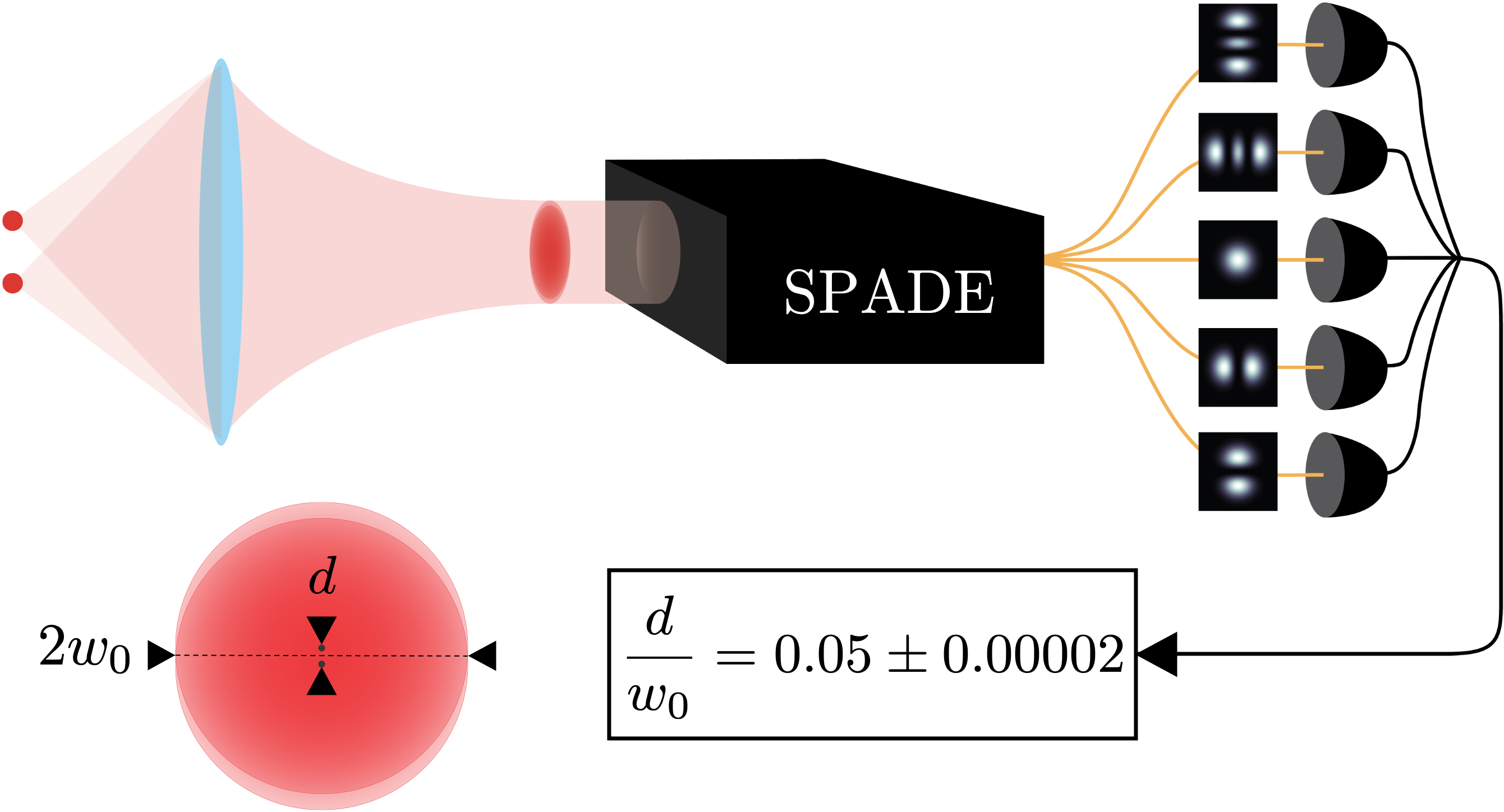}
    \caption{\textbf{Spatial-mode demultiplexing for separation estimation.} SPADE consists in decomposing the incident light over the Hermite-Gaussian mode basis. Measuring the intensity corresponding to each mode was demonstrated to be an optimal measurement for transverse separation estimation. It gives a significant advantage with a gain of some orders of magnitude, compared to direct imaging, on the sensitivity of the estimation of close incoherent sources imaged through a diffraction-limited optical system. The numbers presented in this Figure are typical results obtained with our experimental setup, where $d$ is the transverse separation of two beams in the image plane and $w_0$ is the waist of the beams in the same plane.}
    \label{fig:concept}
\end{figure}

The sensitivity and resolution of optical imaging systems play a crucial role in numerous fields ranging from microscopy to astronomy \cite{labeyrie_introduction_2006,henning_protostars_2014,dekker_resolution_1997,nan_single-molecule_2013,baddeley_biological_2018}. The core challenge, often used as a performance benchmark, is how precisely the separation between two incoherent point sources can be resolved. This key problem dates back to the late 19th century: the technological advances in microscopy and astronomy enabled scientists to observe objects at higher magnifications and with unprecedented levels of detail, but limited by diffraction. Understanding the fundamental principles that govern the behavior of light was essential to improve the resolution and accuracy of instruments. Thus, some of the leading optical physicists of the time turned their attention to this problem: Abbe \cite{abbe_beitrage_1873}, Rayleigh \cite{rayleigh_xxxi_1879} and later Sparrow \cite{sparrow_spectroscopic_1916} proposed criteria based on visual benchmarks and diffraction properties of light. We know today that diffraction alone does not set a fundamental limit, but combined with detectors characteristics and noise sources defines practical boundaries \cite{treps_quantum_2003}. Super-resolution techniques that circumvent the diffraction limit have emerged over the last decades \cite{hell_breaking_1994, betzig_imaging_2006,hell_far-field_2007,dickson_off_1997}. However, these domain-specific techniques are hitherto limited to certain types of microscopy. They require either intricate control over the light source \cite{hell_breaking_1994,hell_far-field_2007} or manipulations of the illuminated sample \cite{betzig_imaging_2006,dickson_off_1997}. Thus, these techniques are incompatible with passive imaging, where one does not control the properties of the light incoming from the scene to be imaged. \\
Passive imaging with spatially resolved intensity measurement, a strategy known as direct imaging (DI) that makes use of high-performance cameras, provides only a limited improvement and prevents substantial advancement beyond the limit imposed by the Rayleigh criterion \cite{shahram_statistical_2006}. However, recently, Tsang et al. \cite{tsang_quantum_2016} approached the historic problem of estimating the separation between two incoherent point sources adopting the framework of quantum metrology. 
They demonstrated that the use of spatial-mode demultiplexing (SPADE) combined with intensity measurements is optimal, in the sense that it saturates the ultimate limit imposed by the laws of physics --the quantum Cramér-Rao bound \cite{helstrom_resolution_1973,giovannetti_advances_2011,barbieri_optical_2022}. SPADE provides a scaling advantage for the minimal resolvable distance compared with DI in an ideal scenario. This advantage is preserved in the presence of experimental noise even if the scaling is degraded \cite{gessner_superresolution_2020,len_resolution_2020,lupo_subwavelength_2020,sorelli_optimal_2021}.
The advantages provided by this metrology-inspired approach have been extended to optical imaging \cite{pushkina_superresolution_2021,bearne_confocal_2021} and other related problems such as discrimination tasks \cite{lu_quantum-optimal_2018, grace_identifying_2022} and multiparameter estimation \cite{rehacek_multiparameter_2017, napoli_towards_2019, tsang_subdiffraction_2017}, also including more general photon statistics \cite{nair_far-field_2016,lupo_ultimate_2016}.\\
Early experiments used interferometric schemes to implement a simplified version of the demultiplexing approach \cite{yang_far-field_2016,paur_achieving_2016,tang_fault-tolerant_2016,tham_beating_2017,zhou_quantum-limited_2019,zanforlin_optical_2022,parniak_beating_2018,wadood_experimental_2021}, emulating the incoherence of the sources and restricting the estimation to short separations by accessing only two modes. Recently, multi-plane light conversion \cite{morizur_programmable_2010} has emerged as a promising technique for estimating separation, enabling a multimodal approach with the potential to reach the ultimate sensitivity at any separation. Two recent experiments explored this approach in different regimes: Boucher et al. \cite{boucher_spatial_2020} with equal-brightness sources and Santamaria et al. \cite{santamaria_spatial-mode-demultiplexing_2023} with a strong brightness imbalance. They demonstrated that this technique is potentially efficient, but did not achieve an ultra-sensitive separation estimation.\\
Here, we implement separation estimation of two incoherent equally-bright sources using spatial-mode demultiplexing over five spatial modes, combined with intensity measurements (see Figure~\ref{fig:concept}). For bright sources we directly measure a sensitivity up to five orders of magnitude beyond the Rayleigh criterion (in practice $\SI{20}{\nano\meter}$ sensitivity with $\SI{1}{\micro\meter}$ accuracy for a $\SI{1}{\milli\meter}$ beam size). For faint sources, we show performances unreachable with even ideal direct imaging (infinite resolution camera, no noise, equivalent losses) and demonstrate $\SI{20}{\micro\meter}$ precision for a $\SI{1}{\milli\meter}$ beam size and approximately 200 measured photons in the selected mode.  Our experiment is the first practical demonstration of passive imaging going significantly beyond the Rayleigh limit, using a simple setup adaptable to standard passive imaging systems and with high-speed performance.

\section*{Experimental setup}
\begin{figure}
    \centering
    \includegraphics[width=0.45\textwidth]{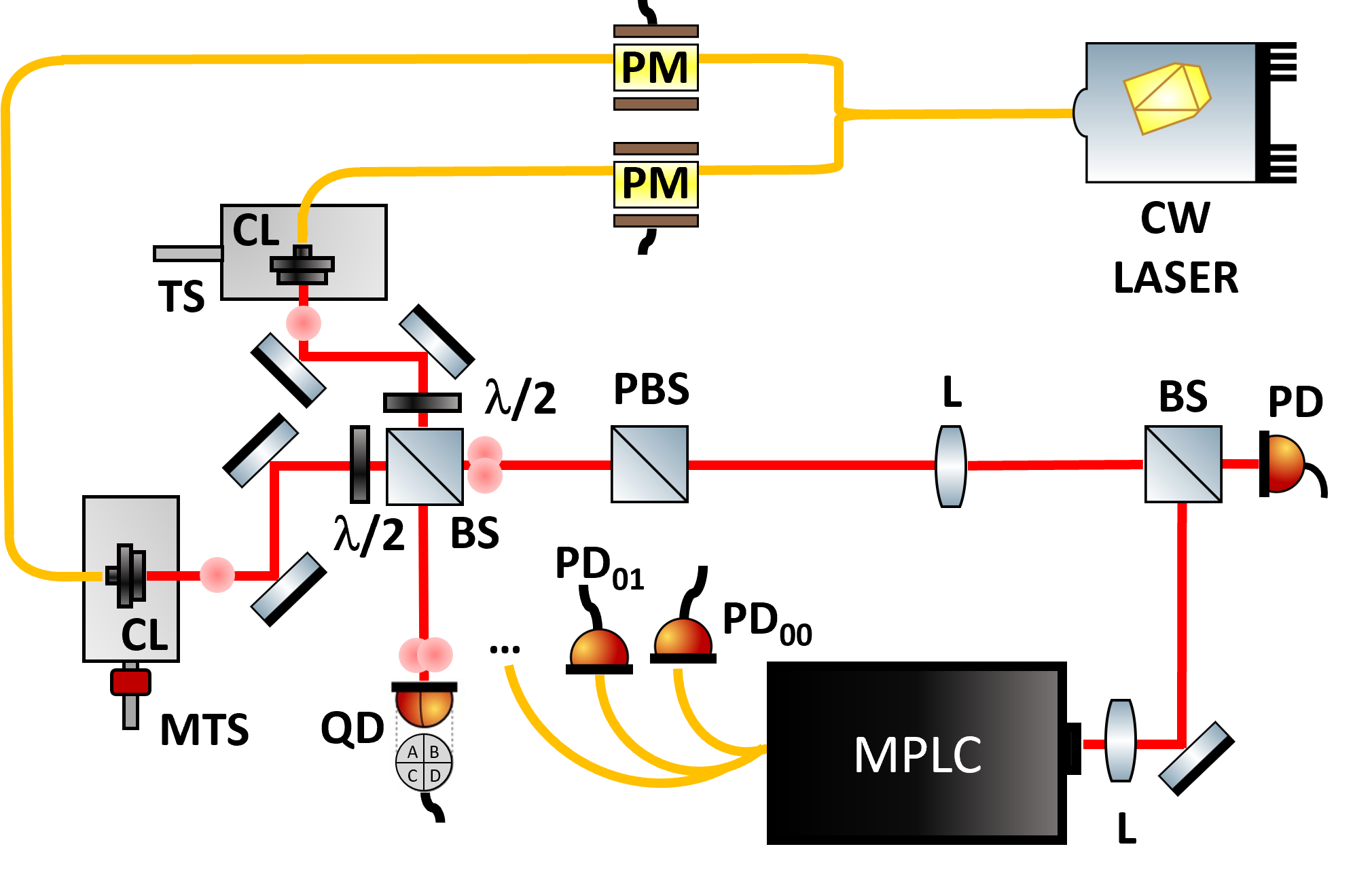}
    \caption{\textbf{Experimental setup.} The two incoherent sources are generated from one continuous-wave fibered laser. The light is split into two paths that are modulated independently with two phase modulators (PM). The beams are then coupled into free space with collimators (CL) fixed either on a translation stage (TS) or on a motorized translation stage (MTS), which are used to set the separation between the two beams. The waist at the output of the collimators is $\SI{1.135}{\milli\meter}$. The multiplexed beam is finally coupled into the multi-plane light converter (MPLC), which intrinsic waist is around $\SI{320}{\micro\meter}$ and the optical powers corresponding to HG modes (HG$_{00}$, HG$_{01}$, HG$_{10}$, HG$_{02}$, HG$_{20}$) are measured with photodetectors (PD$_{00}$, PD$_{01}$...). The reference separation is estimated by determining the position of each beam on a quadrant detector (QD), when the other source is turned off. An external photodiode (PD) is used for normalization in the high-flux regime.}
    \label{fig:setup}
\end{figure}

The experimental setup is detailed in Figure~\ref{fig:setup}. The spatial-mode demultiplexing system is a multi-plane light converter (MPLC, Proteus-C from Cailabs). It decomposes an input light beam on the Hermite-Gaussian (HG) mode basis, each mode being subsequently coupled to a single-mode fiber. It allows for intensity measurements on several HG modes simultaneously. We use five MPLC outputs (out of ten) corresponding to the modes HG$_{00}$, HG$_{01}$, HG$_{10}$, HG$_{02}$ and HG$_{20}$. At the detection stage we use either photodiodes or single-photon avalanche-photodiodes depending on the input light flux.

Two incoherent optical sources are generated as follows: the light from a single fibered-telecom CW laser is split into two paths and goes through independent electro-optical modulators that apply random phases with high frequencies (see Methods). The two incoherent guided modes are free-space coupled by collimators and combined on a beam splitter, thus mimicking the images of two point sources separated by a set separation. They are mounted on independent translation stages so that the transverse separation between the two beams is adjustable. The two beams are Gaussian with similar sizes given by a waist of $w_0 \approx\SI{1.135}{\milli\meter}$. The combined beam is imaged at the input of the MPLC, and mode-matched to its waist $w_{1}\approx\SI{320}{\micro\meter}$.
A photodiode monitors the power stability and half-wave plates combined with a polarizing beam splitter allow to balance the brightness of the two sources.

The measurement device is calibrated using only one source, with a position reference that is a quadrant detector in our case. Note that for single source position estimation the quadrant detector allows for close-to-Cramér-Rao-bound-limited estimation, and is thus a trustable reference \cite{fabre_quantum_2000}. The single beam is aligned and centered on the MPLC using the five HG modes intensities. Usage of all these modes, delivering information on both the centroid and mode-matching \cite{chille_quantum_2015}, is critical for the robustness and repeatability of the procedure. To proceed with calibration, the beam is translated in discrete steps whose position is determined with the quadrant detector (see Methods) and the HG$_{01}$-mode output intensity is measured. This mode carries all the information necessary to estimate small transverse displacements ($\lesssim w_0$) \cite{delaubert_tem_2006}, which is the regime we consider in this article. Thanks to this precise procedure, this calibration  can be done once and for all, and used for every estimation. It becomes a specification of the apparatus that does not need to be checked on a daily basis.

This calibration curve is used to perform parameter estimation. In our case, we make the hypothesis that the scene is composed of two identical incoherent sources and that the centroid is known. This information is used to compute a "two-source" calibration curve from the symmetrisation of the apparatus calibration curve (see Methods). It allows to infer the separation between two sources from the knowledge of the optical power in mode HG$_{01}$. 

Then both optical sources are turned on, the beams are aligned and centered on the MPLC in an identical manner using the five HG output intensities and the quadrant detector. We perform a measurement as follows. First, the two beams are displaced symmetrically by a certain distance, keeping the centroid unchanged (the symmetry is guaranteed by the quadrant photodiode, but in this configuration, this photodiode is unable to deliver any information on the separation -- see Methods). Then the optical power at the output of the HG$_{01}$ fiber is measured over a specific integration time. Finally, the separation is estimated using the "two-source" calibration curve. For each optical setting, this measurement is repeated 200 times in order to evaluate the statistical error of the measurement (the estimated separation is then the average of the 200 estimations, and the error on this value is given by the statistical standard deviation).

 We estimated several separations in two intensity regimes ($3500$ and $10^{13}$ detected photons per integration time).


\section*{Low-flux regime}
\begin{figure}[ht]
\centering
    \includegraphics[width=0.5\textwidth]{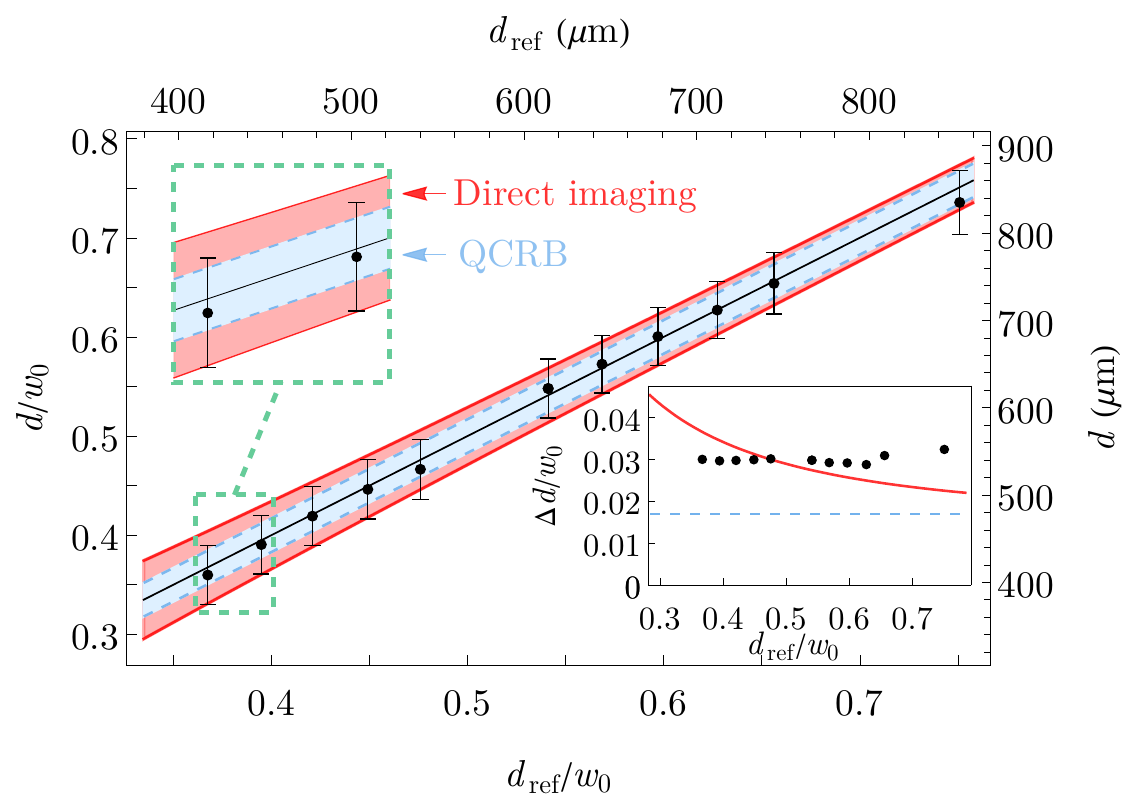}
    \caption{\textbf{Low-flux  measurements.}  The separation estimation is realized with faint sources (3500 photons detected during $\SI{100}{\milli\second}$), using the mean value of the measured intensity corresponding to HG$_{01}$ and the calibration curve. The estimated separations are plotted as a function of the reference separations determined  with the quadrant detector. Both axes are presented with absolute values and values relative to the size of the beam. Error bars due to statistical uncertainty on the reference separation and the estimation, determined with 200 measurements (each during one integration time of $\SI{100}{\milli\second}$), are displayed as well as the unbiased estimation line (black line). The quantum Cramér-Rao bound -- QCRB -- (light blue) and the Cramér-Rao bound for ideal direct imaging (red) are also plotted as shaded areas for comparison. In the inset, we plot the sensitivity of the SPADE measurement as a function of the separation, along with the quantum Cramér-Rao bound (dashed blue line) and the Cramér-Rao bound for perfect direct imaging (red line), calculated for the same number of detected photons.}
    \label{fig:APD}
\end{figure}

 We first present our results for the separation estimation between two faint sources, where the total incident power on the MPLC is around $\SI{50}{\femto\watt}$ during an integration time of $\SI{100}{\milli\second}$ (resulting in 3500 detected photons). In Figure~\ref{fig:APD}, we plot the estimated separation as a function of the reference separation $d_{\mathrm{ref}}$, for separations going from $\SI{400}{\micro\meter}$ up to $\SI{860}{\micro\meter}$. We see a perfect linear trend and agreement between the measurement performed with the MPLC and the reference separation (obtained by measuring independently the position of each source with the quadrant detector (see Methods).

 In order to benchmark the performance of the estimation we compute the statistical standard deviation, as explained in the experimental setup section, and represent it as error bars on the experimental points in Figure~\ref{fig:APD}. We also plot this measurement sensitivity as a function of the separation in the inset of this figure. The measured sensitivity is very close to the quantum Cramér-Rao bound ($\SI{19}{\micro\meter}$ when calculated for the same number of measured photons), the discrepancy arising from the level of dark counts of the detectors. As a matter of comparison, we also compute the classical Cramér-Rao bound for separation estimation with ideal direct imaging, considering infinitely small pixels, no noise, and the same detector quantum efficiency as for our experiment (see Methods). Remarkably, our scheme outperforms this idealised DI setting for small separations ($<\SI{500}{\micro\meter}$). Furthermore, DI requires the measurement of all the photons (3500 in our case), increasing the influence of experimental noise, while modal decomposition allows to route the photons that carry information into a specific output of the MPLC (corresponding to the HG$_{01}$ mode) and perform the detection on only 200 photons with a single detector to deliver an efficient estimator.\\

\section*{High-flux regime}
We consider now bright optical sources, which correspond to an incident power on the MPLC of around $\SI{650}{\micro\watt}$ for an integration time of $\SI{5}{\milli\second}$ (or $10^{13}$ detected photons). Due to the scaling of the sensitivity versus the number of detected photons, we expect a much better sensitivity in this regime. 

\begin{figure*}[ht]
\centering
    \includegraphics[width=0.47\textwidth]{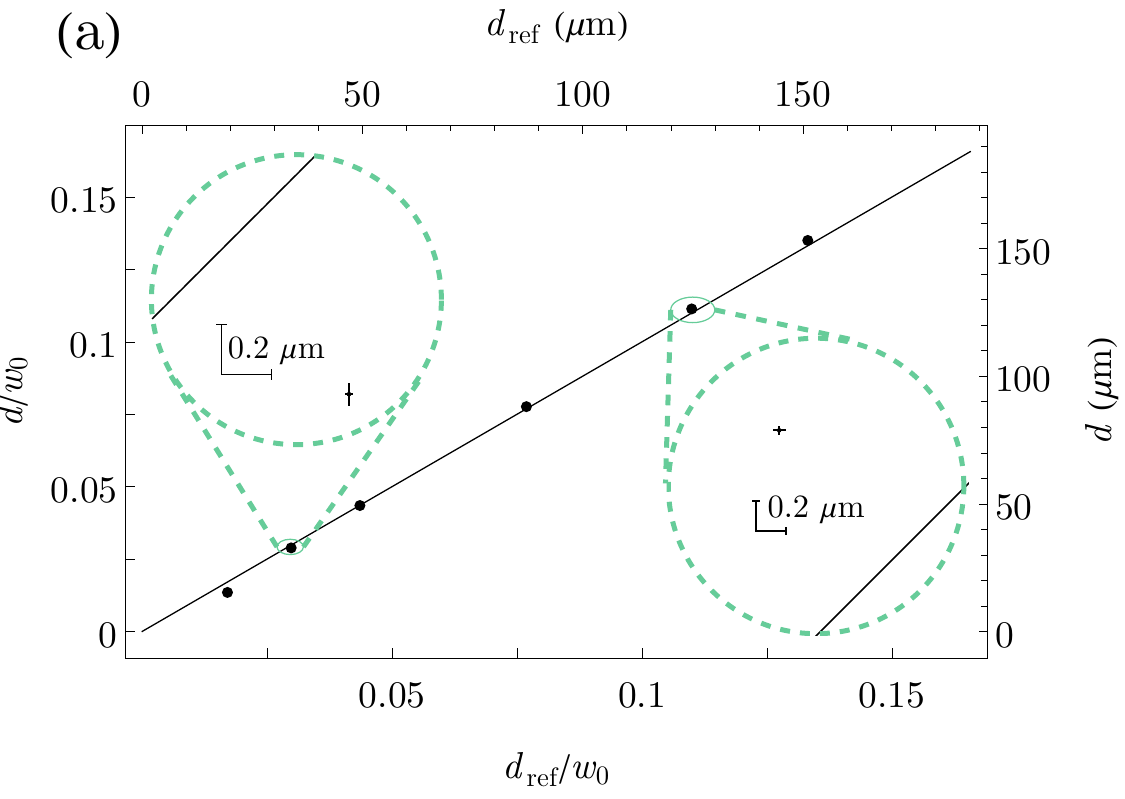}
    \includegraphics[width=0.47\textwidth]{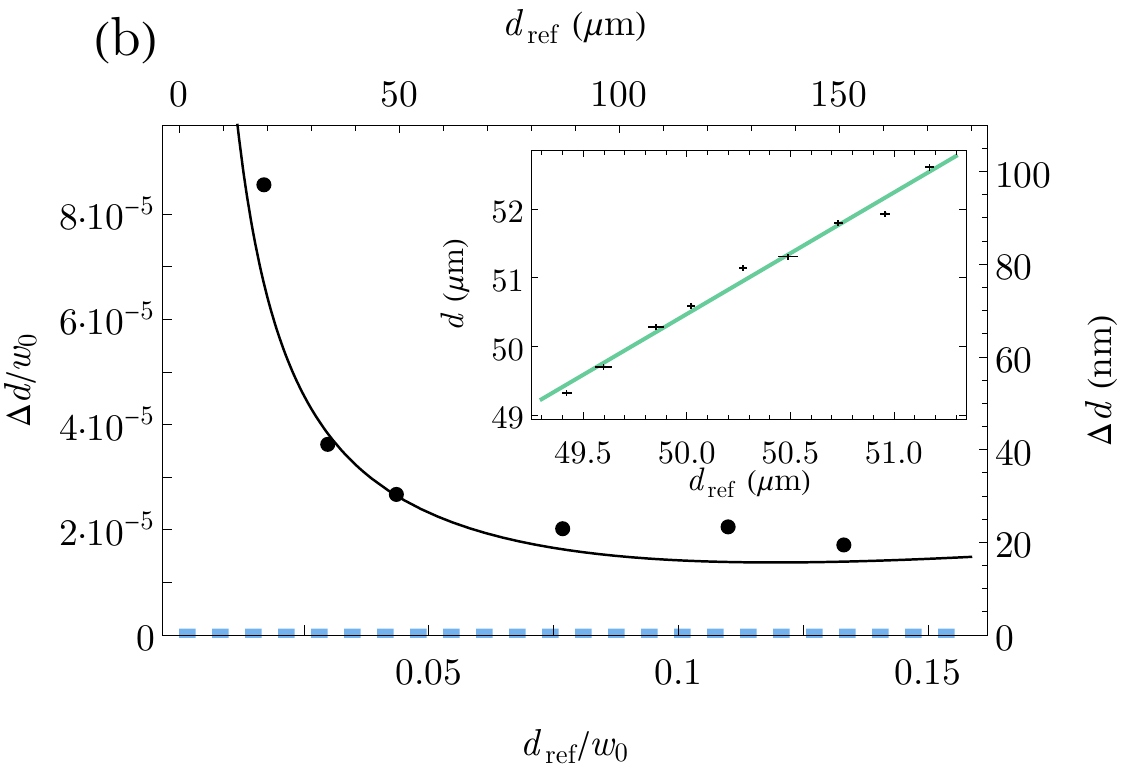}
    \caption{\textbf{High-flux measurements.}  The separation estimation is realized with bright sources (${10^{13}}$ photons detected during $\SI{5}{\milli\second}$), using the mean value of the measured intensity corresponding to HG$_{01}$ and the calibration curve. (a) The estimated separations are plotted as a function of the reference separations determined  with the quadrant detector. Both axes are presented with absolute values and values relative to the size of the beam. Error bars due to statistical uncertainty on the reference separation and the estimation, determined with 200 measurements (each during one integration time of $\SI{5}{\milli\second}$), are displayed as well as the unbiased estimation line (black line). (b) The quantum Cramér-Rao bound (dashed blue line), calculated for the same number of detected photons, and the sensitivity for the SPADE measurement taking into account the detection noise (black line) are also plotted for comparison. In the inset, we plot the estimated separations versus the reference separations when performing a differential measurement. A separation of around $\SI{50}{\micro\meter}$ is fixed, and one source was displaced by several steps, each of approximately $\SI{200}{\nano\meter}$. For each point, 200 measurements (each during one integration time of $\SI{5}{\milli\second}$) were realized to determine the statistical errors. The experimental points follow a linear tendency (green line).}
    \label{fig:Photodiodes}
\end{figure*} 
\noindent

Similarly to the low-flux regime, we plot the estimated separation as a function of the reference separation in Figure~\ref{fig:Photodiodes}(a), for separations ranging from $\SI{20}{\micro\meter}$ to $\SI{160}{\micro\meter}$. Once again, we note a perfect linear agreement between the measured separations and the reference ones. The reported error bars are computed from statistical error estimation. We can see on the inset of Figure~\ref{fig:Photodiodes}(a) that while these error bars are barely visible, the estimated separation deviates on average by approximately $\SI{1}{\micro\meter}$ from the reference separation. This deviation from unbiased estimation limits the accuracy of our system in the high-flux regime. This limitation can be traced back to the small differences between the two sources, because the two-source separation estimator is constructed from the single-source calibration under the hypothesis that both sources are identical. In practice the images of the sources have slightly different spatial shapes, which originates from how we generate them. This deviation amounts for the $\SI{1}{\micro\meter}$ limit (see Supplementary Materials). This is consistent with such an ultra-sensitive apparatus and is not due the detection system itself. In realistic scenarios, like microscopy or astronomy setups,  this accuracy value will depend on the optical imaging device and the type of sources, and can eventually be largely improved.



 To elucidate the full potential of our ultra-sensitive apparatus, we now focus on the statistical errors. To do so, we perform differential measurements: from a scene with a given separation between the two sources, for instance $\SI{50}{\micro\meter}$, only one source is displaced by a series of very small steps, of approximately $\SI{200}{\nano\meter}$ each. At each step, both separation and sensitivity estimation are performed. The results are displayed in the inset of Figure~\ref{fig:Photodiodes}(b), where we plotted the experimental points along with a linear fit as a guide for the eye. We observe in this case statistical errors of about $\SI{20}{\nano\meter}$, and a linear trend consistent with the error bars. The slope of the linear fit is not equal to one, due to the limited accuracy, however, this demonstrates that our apparatus displays the unprecedented ability to distinguish between two scenes with a difference in separation of the order of $\SI{20}{\nano\meter}$. Note that some slight deviations from the estimated separation and the linear fit can be observed, this is to be expected in such an ultra-sensitive measurement where the actual scene is dependent on any mechanical or electronic noise (see Methods).

Finally, we plot in Figure~\ref{fig:Photodiodes}(b) the sensitivity of source separation estimation versus the value of the separation and compare it to theoretical calculations. We demonstrate sensitivities ranging from $\SI{97}{\nano\meter}$ for our shortest separation of $\SI{20}{\micro\meter}$ to as low as $\SI{20}{\nano\meter}$ for larger separations. This corresponds to five orders of magnitude beyond the beam size. This feature is unique to our system, whose practicality is ensured by the single source independent calibration (made possible by the information from multimode MPLC outputs). We compare these values to the quantum Cramér-Rao bound, which is $\SI{0.4}{\nano\meter}$ for ${10^{13}}$ detected photons. The difference is quantitatively reproduced by our theoretical model (see Methods) taking into account the electronic noise of the detection apparatus, demonstrating that this is the limiting factor in our experiment. Note that this sensitivity corresponds to $\approx\SI{5}{\nano\meter}$ at the input plane of the MPLC due to magnification $1/4$ of the mode-matching telescope.

\section*{Conclusion}

We realized a proof-of-principle experiment that demonstrates the practicability of modal decomposition for sub-diffraction separation estimation. We achieved a groundbreaking sensitivity in the estimation of the separation between two incoherent point sources for high and low brightness --beating the diffraction limit by five orders of magnitude and outperforming ideal direct imaging, respectively--, thus setting a new standard in optical resolution. By leveraging the multimode nature of our experiment, we accomplished a robust calibration that led to this unprecedented level of sensitivity. Our system is simple to implement and can be adapted to advanced imaging systems, is fast and requires few detectors thus adaptable to any input light flux. Notably, reducing noise sources --particularly at the detection stage-- could enhance the separation-estimation sensitivity even further, eventually reaching the ultimate quantum limit.\\

Moreover, our singular scheme allows to explore more complex scenes. The source-phase modulation opens a new avenue for the study of tunable coherence situations \cite{wadood_experimental_2021}. In addition, an immediate extension of this work is the exploitation of the MPLC higher-order modes to estimate larger separations \cite{sorelli_moment-based_2021}, but also for multi-parameter estimations \cite{rehacek_optimal_2017,rehacek_multiparameter_2017}, leading to a more complete scheme of passive imaging \cite{pushkina_superresolution_2021}. Finally, given the versatility of our approach and setup, we believe this framework can be further developed for applications in microscopy and astronomy within the near future.\\

\section*{Acknowledgments}
We thank Manuel Gessner, Claude Fabre, and Valentina Parigi for fruitful discussions. We thank Eleni Diamanti for providing two single-photon avalanche photodiodes. This work was partially funded by French ANR under COSMIC project (ANR-19-ASTR0020-01). This work was supported by the European Union’s Horizon 2020 research and innovation programme under the QuantERA programme through the project ApresSF. This work was carried out during the tenure of an ERCIM ‘Alain Bensoussan’ Fellowship Programme.

\bibliography{references}

\newpage
\appendix
\section{Methods}

\textbf{Generation of the two sources} The two incoherent beams are generated from a continuous-wave fibered laser at $\SI{1550}{\nano\meter}$ (Thorlabs SFL1550P, coherence length $\SI{6}{\kilo\meter}$) that is split into two paths. Each path is then phase modulated independently using electro-optical phase modulators (iXblue MPX-LN-0.1). These modulators are driven with independent FPGA hardware boards (Red Pitaya STEMlab 125-14) generating voltages uniformly distributed between -1V and 1V, which are then amplified to reach the voltage amplitude needed for covering [$-\pi$,$\pi$], at $\SI{4}{\mega\hertz}$. The incoherence of the sources is assessed by the vanishing of the interference fringes (see Supplementary Materials). This configuration allows to have access to a tunable coherence between the two sources if needed for future experiments.
\bigbreak
\noindent
\textbf{Detection} The spatial-mode demultiplexer (Cailabs Proteus-C) has ten outputs but we only use five of them (corresponding to HG$_{00}$, HG$_{01}$, HG$_{10}$, HG$_{02}$ and HG$_{20}$) as we restrict ourselves to the estimation of short separations, i.e. shorter than the beam radius $w_{0}$. The MPLC presents -3.9 dB (59\%) of losses. At the fiber outputs, photodetection is realized either with switchable-gain photodiodes (Thorlabs PDA50B2, 85\% responsivity) --using a gain of 60 dB (bandwidth $\SI{630}{Hz}$) for the HG$_{01}$ output-- or with single-photon avalanche-photodiodes (IDQuantique ID230, 25\% quantum efficiency). \\
For the high-flux measurements (around $\SI{100}{\micro\watt}$ entering the MPLC), the integration time used for one measurement is $\SI{5}{\milli\second}$. This detection timescale ensures that we average enough to measure incoherent beams.  We normalize the optical powers at the outputs of the MPLC with about 10\% of the total incident light, measured with an external photodiode (Thorlabs PDA20CS-EC). Data are acquired with two oscilloscopes (Tektronix 4 series MSO44) at a sample rate of $\SI{25}{\kilo\hertz}$ for a total measurement time of $\SI{10}{\second}$. Due to the high sensitivity of high-flux measurements, we are extremely sensitive to external noises such as mechanical noise, air flow and fluctuations in detector offsets. In order to dissociate from these noise sources, we chose to determine the statistical errors over only $\SI{1}{\second}$, which corresponds to $\si{200}$ consecutive points (see Supplementary Materials). \\
For the low-flux regime (incident optical power around $\SI{50}{\femto\watt}$), the integration time used for one measurement is $\SI{100}{\milli\second}$ to collect enough data for statistical analysis.  In this case, as we have only two single-photon detectors, we normalize the optical power of the HG$_{01}$ mode with the one measured at the output of the HG$_{00}$ mode. The total measurement time with faint sources is $\SI{20}{\second}$, resulting in 200 points.
\bigbreak
\noindent
\textbf{Reference separation} We estimate the displacement of each source using a quadrant detector (Thorlabs PDQ30C) signal. This detector is composed of two photodiodes and the difference between their output currents gives access to the displacement of the beam. Remarkably, this means that the quadrant detector is blind in the presence of two symmetric sources and it cannot be used for estimating a separation. During the calibration, we use it to determine the displacement of the source with respect to the centroid of the MPLC --on which the quadrant detector is aligned. Then, during the measurement, we measure the displacement of each source while the other is turned off and we then extract the reference separation $d_{\mathrm{ref}}$ and the associated statistical error bars ranging from $\SI{20}{\nano\meter}$ to $\SI{50}{\nano\meter}$.
\bigbreak
\noindent
\textbf{Alignment procedure} To ensure high accuracy and sensitivity, it is critical to align the two beams identically. We use bright beams to this end. First, each beam is roughly centered on the MPLC by maximizing the signal in the HG$_{00}$ mode. Then, both beams are mode-matched on the MPLC: the focal lengths and the positions of the lenses are chosen to minimize the signals in the HG$_{02}$ and HG$_{20}$ modes. Finally, we minimize the HG$_{01}$ and HG$_{10}$ signals with a pair of steering mirrors --per beam-- in order to improve sensitivity on the position and tilt of the beams at the input of the MPLC. The quadrant detector is used to ensure that they are aligned on the same center point.
\bigbreak
\noindent
\textbf{Calibration and estimation} After the two beams are aligned, we use only one to perform the calibration of the setup ensuring that calibration and source-separation measurement are independent. The beam is displaced with respect to the MPLC center with a motorized translation stage. We cover a position range from $\SI{-100}{\micro\meter}$ to $\SI{100}{\micro\meter}$ around the MPLC center, in steps of $\SI{6}{\micro\meter}$. The optical power corresponding to the HG$_{01}$ mode is measured during $\SI{10}{\second}$ --with a photodiode or a single-photon avalanche photodiode depending on the flux-- for every position. Its mean value is normalized with the mean value of the optical power measured by the external photodiode and in the HG$_{00}$ mode for bright and faint sources, respectively. The calibration curve for one source $f(x)$ is obtained from a polynomial fit to the experimental points.\\
For each set separation, we measure the normalizing optical power $p_{1,2}$ and the quadrant-detector signal $x_{1,2}$ corresponding to each source. We can then determine the reference separation value ${d_{\mathrm{ref}}= \mid x_1-x_2\mid}$ and the two-source calibration curve for the separation \\ ${g(d=2x)=\frac{p_1}{p_1+p_2}f(x) + \frac{p_2}{p_1+p_2} f(-x)}$. This allows to take into account relative power changes due to polarization fluctuations in the independent fiber paths.\\
Finally, to extract the estimated separation and its associated Finally, to estimate the separation, we average measured optical power in the HG$_{01}$ mode during the integration time ($\SI{5}{\milli\second}$ for the high flux or $\SI{100}{\milli\second}$ for the low flux) obtaining $p^i_{01}$. We normalize the result with total power $p^i_{\mathrm{tot}}$, measured either from the HG$_{00}$ mode  for low flux or from the external photodiode for high flux. We repeat the procedure 200 times, the estimation $d$ and its sensitivity $\Delta d$ is then obtained by determining the mean value and the standard deviation over all the $d_i$, extracted by inverting the two-source calibration curve for each bin $i$. We assess the accuracy of our estimation by comparing it to the reference separation $d_{\mathrm{ref}}$ obtained with the quadrature detector.
\bigbreak
\noindent
\textbf{Theoretical sensitivity}
We estimate the sensitivity of ideal direct imaging using the Fisher information. Considering continuous noiseless detection of the intensity distribution the Fisher information per unit time takes the form
\begin{equation}
    F = \frac{1}{E_\text{ph}} \int \frac{1}{I(\vec r)} \left( \frac {\partial I(\vec r)}{\partial d}\right)^2 d \vec r,
\end{equation}
where $E_\text{ph}$ is the energy of the photon. 
The intensity distribution $I(\vec r)$ for two incoherent sources equals 
$I(x,y)=P_1~u^2(x+d/2,y)+P_2~u^2(x-d/2,y)$, where $P_{1,2}$ are the powers of the sources and $u(x,y)$ is a Gaussian mode profile. Then the smallest possible variance of an estimator of the separation between the sources is given by the Cramér-Rao bound $\Delta^2 d \geq 1/(F~t_\text{int})$, where $t_\text{int}$ is an integration time. This bound is shown in Figure~\ref{fig:APD} in red.

The quantum Cramér-Rao bound (Figure~\ref{fig:APD} and Figure~\ref{fig:Photodiodes}(b) in blue) is given by the inequality \cite{tsang_quantum_2016} $ \Delta^2 d \geq w_0^2/N$, where $N=(P_1+P_2)~t_\text{int}/E_\text{ph}$ is the total number of detected photons. 

To estimate the sensitivity of mode demultiplexing (Figure~\ref{fig:Photodiodes}(b), black line) we use the error propagation method. Since we build our estimator based on the measured intensity $I_{01}$ of the mode  HG$_{01}$, its variance equals 
\begin{equation}
\Delta^2 d = \frac{\Delta^2 I_{01}}{\left(\partial I_{01} / \partial d \right)^2 }.
\end{equation}
The intensity noise $\Delta^2 I_{01}$ consists of the shot noise and the detection noise. The latter is dominant in the high-flux regime when intensity is detected with photodiodes. The detection noise is estimated from the offset measurements (without incident light). 
The intensity of HG$_{01}$ mode is given by \cite{sorelli_optimal_2021}
\begin{equation}
     I_{01}=(P_1+P_2) \frac{d^2}{4w_0^2} e^{-\frac{d^2}{4 w_0^2}}.
\end{equation}
In practice measurement modes are slightly different from Hermite-Gaussian, which is often referred to as cross-talk between modes. However,  analysis shows that the effect of the cross-talk in our case is negligible compared to the detection noise.

\section{Calibration}
\textbf{One-source calibration.} One beam is displaced, using a motorized translation stage, from $\SI{-100}{\micro\meter}$ to $\SI{100}{\micro\meter}$, by steps of $\SI{6}{\micro\meter}$, with respect to the MPLC center. The optical power corresponding to the HG$_{01}$ mode is measured for every position, for $\SI{10}{\second}$, and its mean value is normalized with the mean value of the optical power measured by the external photodiode or in the HG$_{00}$ for bright and faint sources respectively. The experimental points are then fit with a polynomial function which is the calibration curve for one source $f(x)$.\\

\begin{figure}
    \centering
    \includegraphics[width=0.4\textwidth]
    {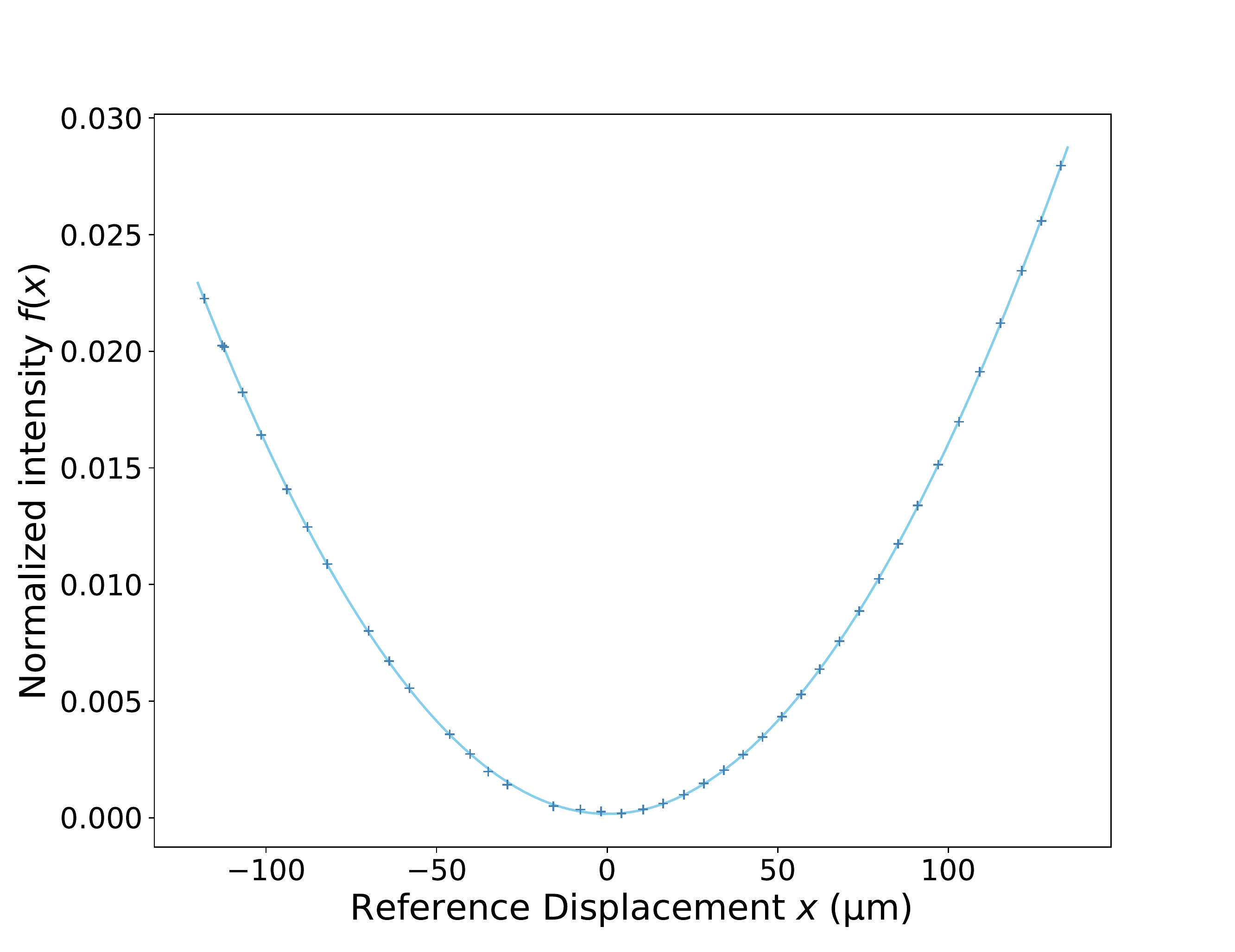}
    \caption{Calibration curve for one displaced beam. The light blue line is a fit of the experimental points with a 6th-order polynomial function.}
    \label{fig:CalibrationDisplacement}
\end{figure}

\begin{figure}
    \centering
    \includegraphics[width=0.4\textwidth]{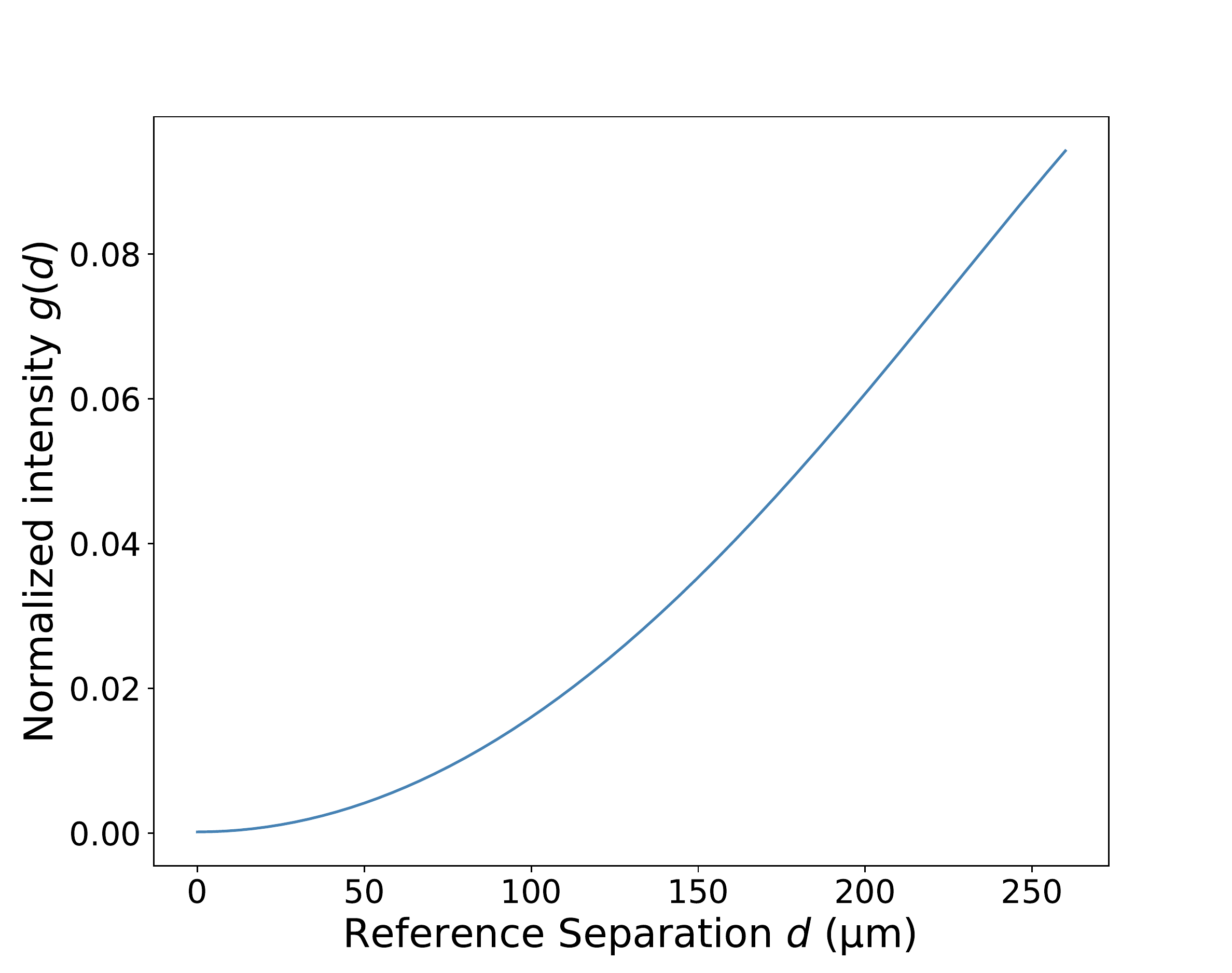}
    \caption{Calibration curve for the separation between two beams. This curve is obtained by symmetrizing the light polynomial function from the one-source calibration curve.}
    \label{fig:CalibrationTwoSources}
\end{figure}

\textbf{Two-source calibration curve.} For each separation, we measure the normalizing optical power $p_{1,2}$, as well as the quadrant detector signal $x_{1,2}$, for each source. We can then determine the reference separation value ${d_{\mathrm{ref}}= \mid x_1-x_2 \mid}$ and the symmetrized calibration curve for the separation between two sources ${g(d=2x)=\frac{p_1}{p_1+p_2}f(x) + \frac{p_2}{p_1+p_2} f(-x)}$. This allows to take into account relative power changes due to polarization fluctuations in the independent fiber paths.
\bigbreak
\noindent
\textbf{Separation estimation.} Finally, to estimate the separation, we average measured optical power in the HG$_{01}$ mode during the integration time ($\SI{5}{\milli\second}$ for the high flux or $\SI{100}{\milli\second}$ for the low flux) obtaining $p^i_{01}$. We normalize the result with total power $p^i_{\mathrm{tot}}$, measured either from the HG$_{00}$ mode  for low flux or from the external photodiode for high flux. To estimate the separation $d$ we solve the equation
\begin{equation}
    \frac{p^i_{01}}{p^i_{\mathrm{tot}}}=g(d_i).
\end{equation}
We repeat the procedure 200 times, the estimation $d$ and its sensitivity $\Delta d$ is then obtained by determining the mean value and the standard deviation over all the $d_i$, and this can be compared to the reference separation $d_{\mathrm{ref}}$.
\vspace{1cm}

\section{Indistinguishability of the two sources}
\begin{figure}
    \centering
    \includegraphics[width=0.45\textwidth]{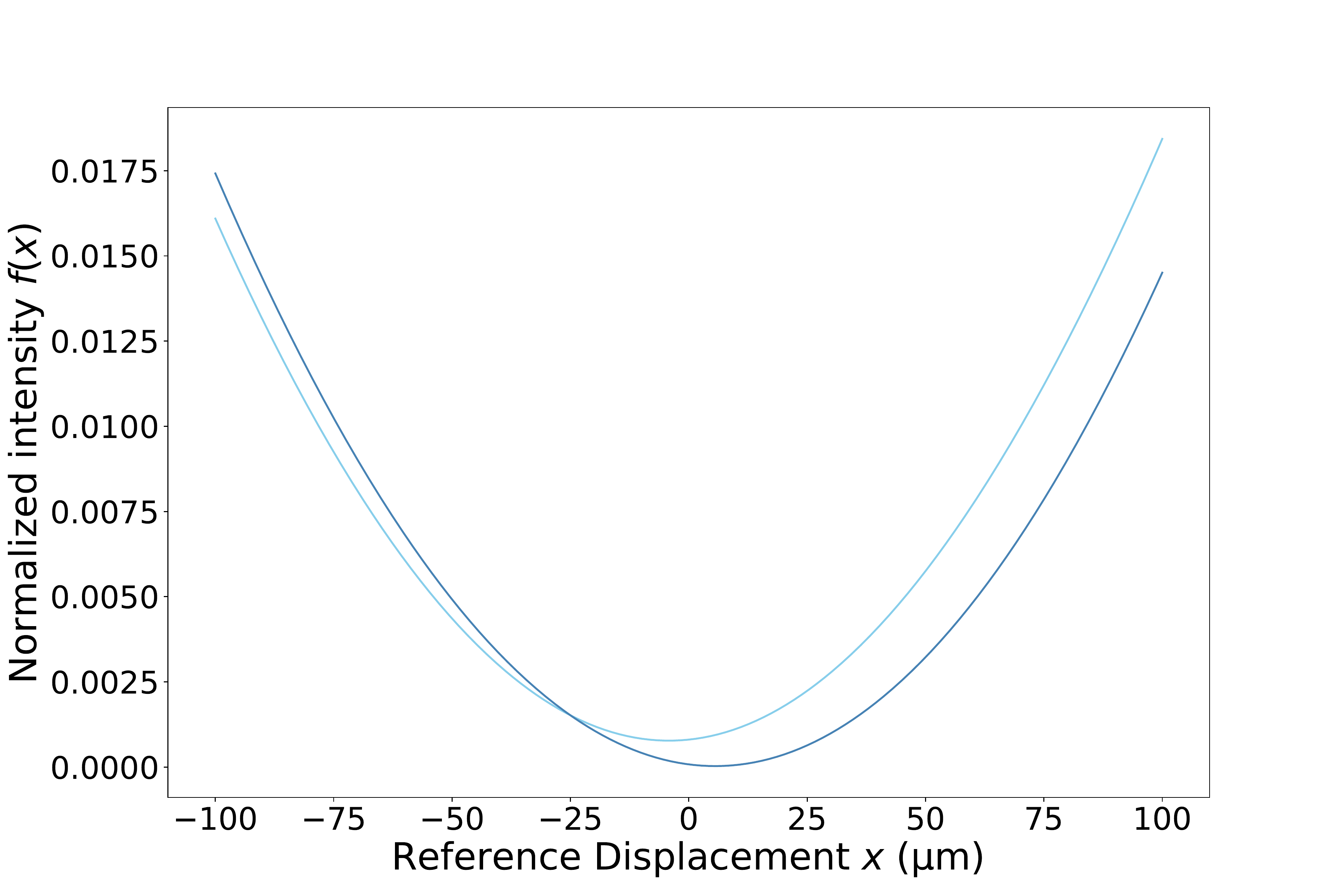}
    \caption{Calibration curves of the two sources for the mode HG$_{01}$.}
    \label{fig:comparaison}
\end{figure}

The central assumption of the separation estimation we implemented is that the two beams are identical. To test this assumption, we performed a calibration for each source. Figure~\ref{fig:comparaison} presents typical calibration curves for the HG$_{01}$ mode of each beam. 

We observe that the two curves are slightly shifted horizontally and vertically from each other. This originates from different tilts and positions of the beams at the input of the MPLC, as well as from different mode shapes, since, in our setup, the two modes are prepared independently. These disparities are the main source of systematic errors. Although this has a small influence on the estimation for symmetric situations, it results in more striking effects when performing differential measurements. Indeed, since one beam remains fixed, this asymmetry in position, combined with the differences in the calibration curves, leads to variations in the slope of the linear tendency (inset plot in Figure 4(b)), depending on the range of separations that are considered. Note, that the trend always remains linear and measurement results stay incredibly sensitive to small displacements of any source.

\section{Estimation in the high-flux regime: temporal signals}
\begin{figure}
    \centering
    \includegraphics[width=0.45\textwidth]{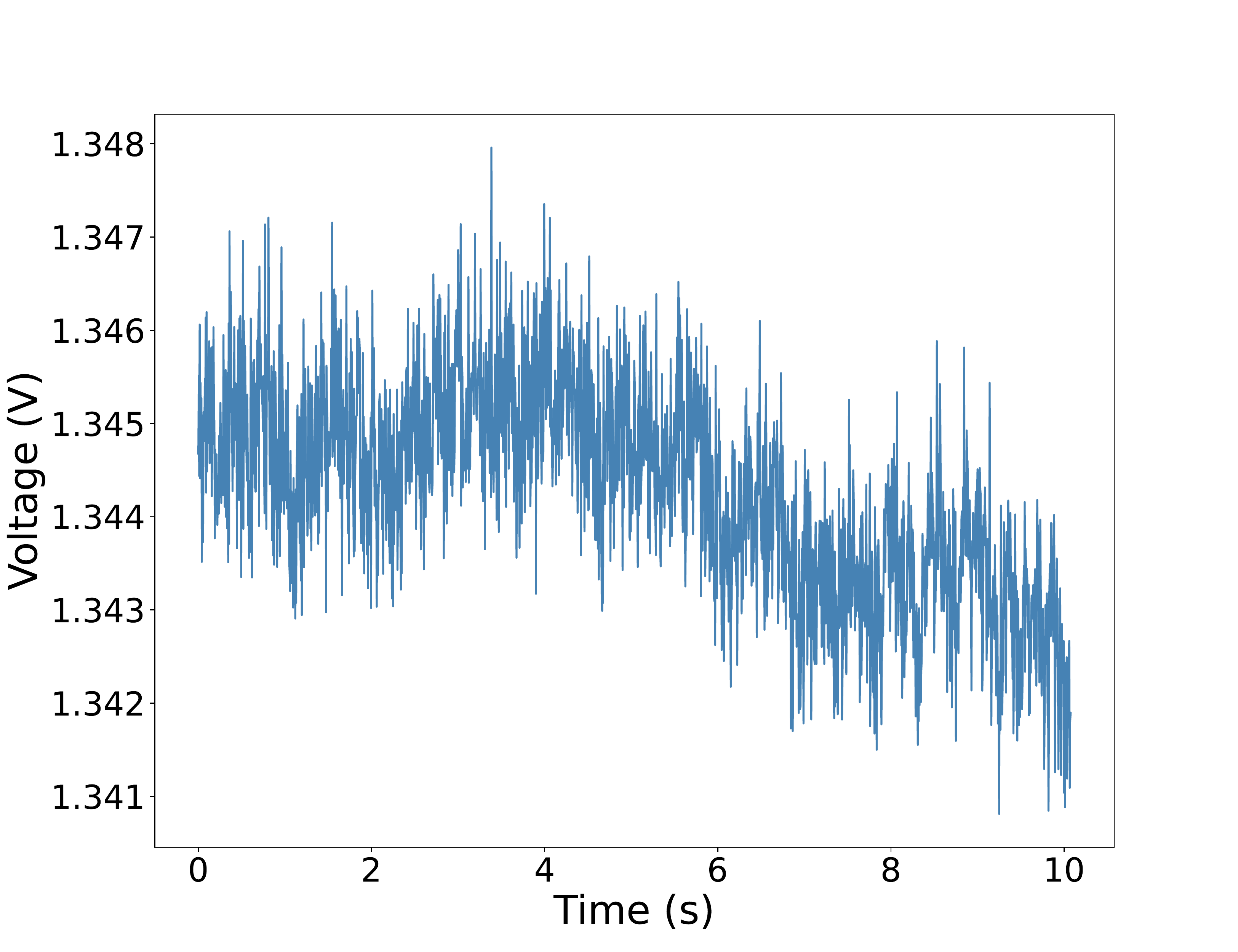}
    \caption{Temporal signal at the output of the HG$_{01}$ mode, acquired with a photodiode (Thorlabs PDA50B2, gain 60dB, integration time $\SI{5}{\milli\second}$) for a transverse separation of around $\SI{120}{\micro\meter}$ between the two beams. }
    \label{fig:temporal}
\end{figure}

As explained in the Methods section, the total measurement time for the high-flux measurements is $\SI{10}{\second}$. Figure~\ref{fig:temporal} presents a typical temporal signal acquired from the photodiode of the HG$_{01}$ output, here for a transverse separation between the two beams of $\SI{120}{\micro\meter}$.

Fluctuations arise from offsets variations, air flows and imperfect incoherence between the two sources, resulting in an increase in the standard deviation of the signal. We chose to reduce the total measurement time to $\SI{1}{\second}$, with an integration time of $\SI{5}{\milli\second}$, to reduce the contributions of these low-frequency noise sources while keeping a sufficient number of points for the statistical analysis.

\section{Incoherent sources}
\begin{figure}
    \centering
    \includegraphics[width=0.45\textwidth]{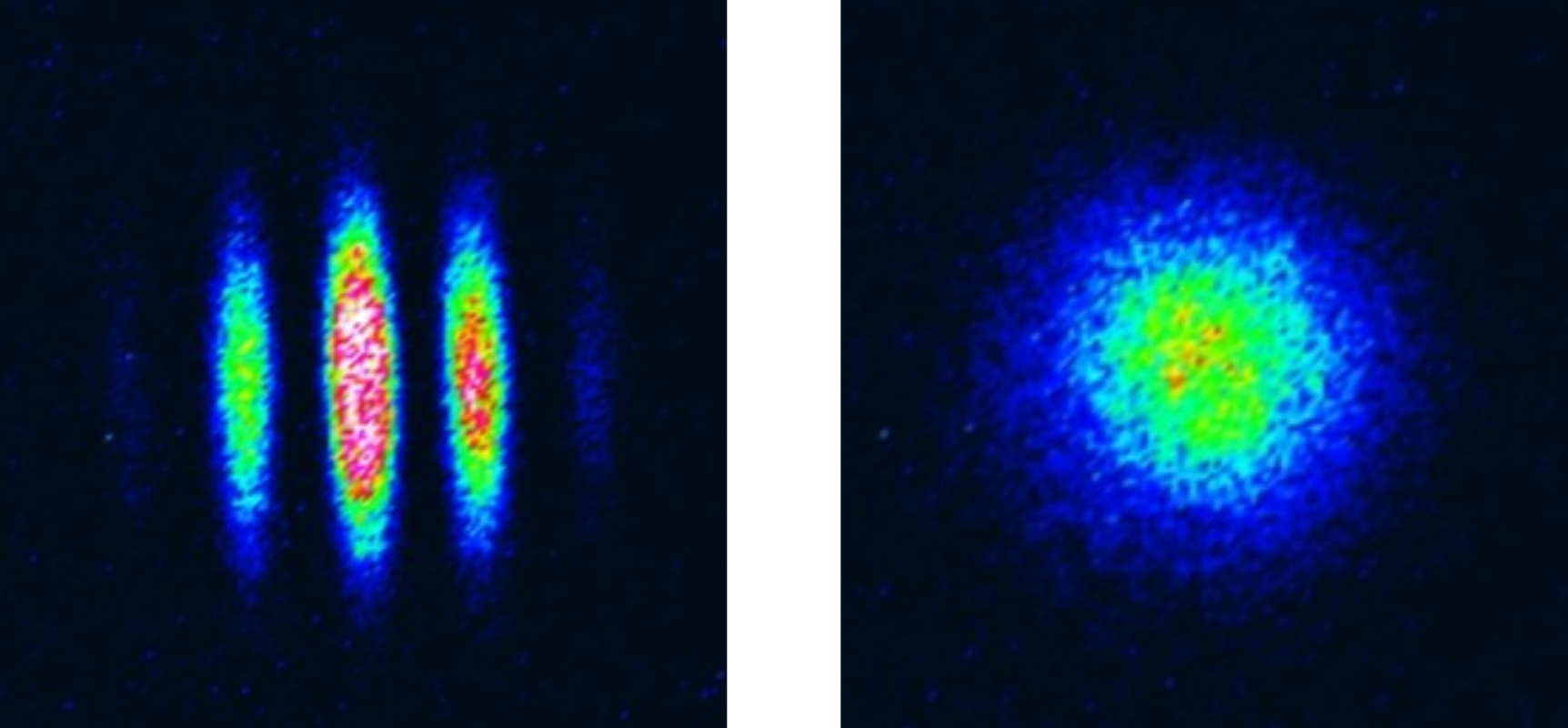}
    \caption{Images of the two beams after mixing on a beam splitter with a CCD camera, without modulation (left) and with a random phase-modulation uniformly distributed between $-\pi$ and $\pi$ (right). }
    \label{fig:incoherence}
\end{figure}

\begin{figure}
    \centering
    \includegraphics[width=0.45\textwidth]{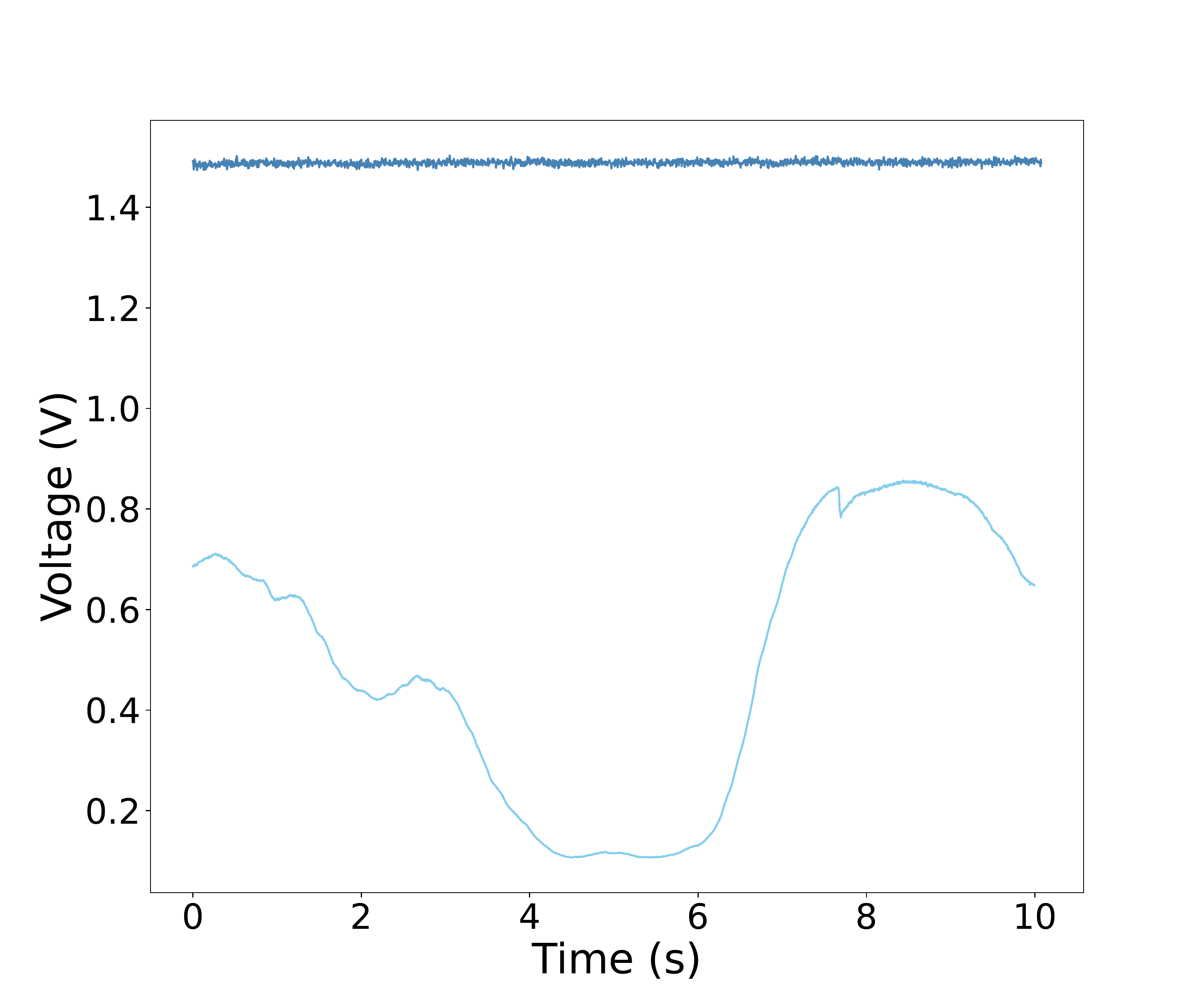}
    \caption{Temporal signals of the two beams after mixing on a beam splitter without (light blue) and with (dark blue) modulation, acquired with a photodiode (Thorlabs PDA50B2, gain 20dB, integration time $\SI{5}{\milli\second}$).}
    \label{fig:incoherence-temporal}
\end{figure}

We generate two incoherent sources from a single laser using two independent phase modulators (see Methods). The vanishing of the interference patterns assesses the incoherence of the beams. Figure~\ref{fig:incoherence} displays images of the two beams after mixing on a beam splitter, without the modulation and with the random modulation, uniformly distributed between $-\pi$ and $\pi$ and with a 4 MHz bandwidth. The images are acquired with a CCD camera (Gentec Beamage-3.0-IR, bandwidth $\SI{8}{\hertz}$).
Figure~\ref{fig:incoherence-temporal} shows temporal signals acquired with a photodiode (Thorlabs PDA50B2, gain 20dB, integration time $\SI{5}{\milli\second}$) after the beam splitter, without (light blue) and with (dark blue) modulation.

\end{document}